\documentclass[preprint2]{aastex}

\usepackage{natbib}
\slugcomment{Accepted for publication in ApJ}

\shorttitle{Spectral Line Survey toward a Molecular Cloud in IC10}
\shortauthors{Nishimura et al.}

\begin{document}

\title{Spectral Line Survey 
       toward a Molecular Cloud in IC10}

\author{Yuri Nishimura\altaffilmark{1},
        Takashi Shimonishi\altaffilmark{2,3},
        Yoshimasa Watanabe\altaffilmark{1},
        Nami Sakai\altaffilmark{4},\\
        Yuri Aikawa\altaffilmark{5},
        Akiko Kawamura\altaffilmark{6},
        and 
        Satoshi Yamamoto\altaffilmark{1}}

\altaffiltext{1}{Department of Physics, the University of Tokyo, 
                 7-3-1, Hongo, Bunkyo-ku, Tokyo, 113-0033, Japan}
\altaffiltext{2}{Frontier Research Institute for 
                 Interdisciplinary Sciences, Tohoku University, 
                 Aramakiazaaoba 6-3, Aoba-ku, Sendai, Miyagi, 
                 980-8578, Japan}
\altaffiltext{3}{Astronomical Institute, Tohoku University, 
                 Aramakiazaaoba 6-3, Aoba-ku, Sendai, Miyagi, 
                 980-8578, Japan}
\altaffiltext{4}{RIKEN, 2-1 Hirosawa, Wako, Saitama 351-0198, Japan}
\altaffiltext{5}{Center for Computational Sciences, The University of Tsukuba, 
                 1-1-1, Tennodai, Tsukuba, Ibaraki 305-8577, Japan}
\altaffiltext{6}{National Astronomical Observatory of Japan, 
                 Osawa, Mitaka, Tokyo, 181-8588, Japan}

\begin{abstract}
We have conducted a spectral line survey observation in the 3 mm band 
toward the low-metallicity dwarf galaxy IC10 
with the 45 m radio telescope of Nobeyama Radio Observatory 
to explore its chemical composition at a molecular-cloud scale ($\sim80$ pc).  
The CS, SO, CCH, HCN, HCO$^+$, and HNC lines are detected for the first time 
in this galaxy in addition to the CO and $^{13}$CO lines, 
while $c$-C$_3$H$_2$, CH$_3$OH, CN, C$^{18}$O, and N$_2$H$^+$ lines are not detected.
The spectral intensity pattern is found to be similar to those observed toward molecular clouds 
in the Large Magellanic Cloud, whose metallicity is as low as IC10.  
Nitrogen-bearing species are deficient in comparison with the Galactic molecular clouds 
due to a lower elemental abundance of nitrogen.  
CCH is abundant in comparison with Galactic translucent clouds, 
whereas CH$_3$OH may be deficient.  
These characteristic trends for CCH and CH$_3$OH are also seen in the LMC, 
and seem to originate from photodissociation regions 
more extended in peripheries of molecular clouds due to the lower metallicity condition.  
\end{abstract}


\section{Introduction}
The Local Group of galaxies comprises 50 or more galaxies, 
a majority of which are low-metallicity dwarf ones.  
Observations of such low-metallicity galaxies 
provide us with a clue to an understanding of chemical characteristics 
of distant galaxies in the early universe, which are in a metal-poor environment.  
It is generally thought that metallicity has a strong impact 
on chemical composition of molecular clouds \citep[e.g.,][]{vandishoeck1988photodissociation, millar1990chemical}.  
In a metal-poor environment, not only are metal-bearing species less abundant, 
but production and destruction processes of molecules are expected to be different 
from those in a metal-rich environment.  
For example, photodissociation and photoionization are more effective 
in molecular clouds in low-metallicity conditions 
owing to the lower abundance of dust grains \citep[e.g.,][]{millar1990chemical}.  
Thus, detailed characterization of the chemical feature in low-metallicity galaxies 
is of fundamental importance in astrochemistry and astrophysics.  

However, radio-astronomical observational studies on chemical composition 
of low-metallicity galaxies have so far been very sparse.  
Molecular line emission from low-metallicity galaxies is generally faint, and hence, 
observations have almost been limited to CO and its isotopologues.  
A notable exception is the Large Magellanic Cloud (LMC).  
Because of its proximity to the Sun 
($d=49.97\pm1.11$ kpc \citep{pietrzynski2013eclipsing}), 
molecular line observations are carried out toward active star-forming regions
\citep[e.g.,][]{chin1997molecular, heikkila1999molecular, wang2009abundances, paron2014aste}.  
Thanks to recent improvement of observing instruments, 
spectral lines of less abundant species are now observable in external galaxies 
with a reasonable observation time, 
which enable us to study their chemical compositions 
\citep[e.g.,][]{martin2006millimeter, aladro2013molecular, watanabe2014spectral}.  
Such a situation is also true even for low-metallicity galaxies.  

We recently conducted spectral line observations 
of molecular clouds in the LMC with the Mopra 22 m telescope 
as the initial step to characterize molecular-cloud-scale chemical composition 
characteristic to low-metallicity condition \citep{nishimura2016spectral}.  
We observed seven molecular clouds, which have different star-formation activities; 
two are quiescent molecular clouds which are not associated with 
infrared counterparts detected with \emph{AKARI} or \emph{Spitzer}, 
three are molecular clouds associated with high-mass star formation, 
and two are active star-forming clouds with the extended H II region.  
The difference of star-formation activities is reflected in infrared fluxes; 
their 8 $\mu$m fluxes vary from 26.8 to 946.8 mJy arcsec$^{-2}$ 
\citep[38 arcsec beam averaged, \textit{Spitzer}/IRAC, ][]{meixner2006spitzer}.
We found that the chemical compositions of the 7 sources 
are similar to each other regardless of the large difference of star-formation activities.  
Since the beam size of the Mopra corresponds to the spatial resolution of 10 pc at the LMC distance, 
the major contribution to the observed spectra is rather diffuse molecular gas 
extending over the molecular clouds.  
The contribution from the star-forming cores seems to be smeared out in the observations.  
In other words, the observed chemical composition 
is characteristic to molecular clouds in the LMC.  

By comparing the results with these observed in the metal-rich environment 
such as our Galaxy and the spiral arm of M51, 
the following two characteristics of the LMC spectra are found \citep{nishimura2016spectral}.  
(1) Nitrogen-bearing species (HCN, HNC, N$_2$H$^+$, HNCO, CN) are obviously faint in the LMC, 
which originates from the low elemental abundance of nitrogen.  
(2) The lines of CCH are relatively brighter and 
the lines of CH$_3$OH are much weaker in the LMC than those in the spiral arm of M51.  
Moreover, the CCH/HCO$^+$ ratios in the LMC are higher than 
those found in the nearby translucent clouds in our Galaxy.  
This trend would originate from the stronger UV radiation field 
owing to the lower abundance of dust grains.  

We expect that these characteristics of the LMC can generally be found 
in other low-metallicity galaxies.  
To examine this prediction, we observed another low-metallicity galaxy IC10. 
IC10 is located at the distance of $\sim$ 950 kpc \citep{hunter2001stellar}, 
and its metallicity is lower by factor 5 than in Solar neighborhood \citep{garnett1990nitrogen}.  
The star formation rate estimated from H$\alpha$ 
\citep[0.2$M_{\odot}$ yr $^{-1}$;][]{gildepaz2003palomar} and 
the association of the large number of Wolf-Rayet stars \citep[$\sim100$;][]{massey2002wolf} 
imply active star formation in IC10.  
IC10 has been studied in C II (158 $\mu$m) \citep{madden1997c}, 
C I ($^3P_1 \rightarrow ^3P_0$) \citep{bolatto2000submillimeter}, 
and multi-transitions of CO 
\citep[e.g.,][]{petitpas1998effect, bolatto2000submillimeter, leroy2006molecular}.  
However, molecular line observations except for CO and its isotopologues have not been reported.  

\section{Observations}
The observations were carried out with 
the 45 m radio telescope at Nobeyama Radio Observatory (NRO) 
in February and March in 2015.  
We observed the most part of the frequency range from 84 to 116 GHz. 
We did not observe the frequency range of $91-91.5$, $95.5-96$, $100-103.5$, and $107.5-108$ GHz, 
because no spectral lines were detected in the LMC sources \citep{nishimura2016spectral}. 
The half-power beam width (HPBW) of the telescope is 
$20.4''$, $16.6''$ and $15.3''$ at 86, 110 and 115 GHz, respectively.  
They correspond to the spatial scale from 71.7 to 98.8 pc 
at the IC10 distance ($d=950$ kpc). 
The telescope pointing was checked by observing a nearby SiO maser source (Y Cas) every hour, 
and the pointing accuracy was maintained to be better than $5''$.  
We observed two orthogonal polarization signals simultaneously 
by using the SIS mixer receiver (TZ1), whose system temperatures ranged from 120 to 280 K.    
The backend is the autocorrelator SAM45.  
The frequency resolution and bandwidth are 488.24 kHz and 1600 MHz, respectively.  
We binded 2 successive channels of SAM45 in the analysis 
to improve the signal-to-noize ratio.  
The resultant velocity resolution is 3.25 km s$^{-1}$ at 90 GHz.
The line intensity was calibrated by the chopper wheel method, 
and a typical calibration accuracy is 20\%.  
The antenna temperature is divided by the main beam efficiency of 
0.49 at 86 GHz, 0.42 at 100 GHz, and 0.40 at 115 GHz 
to obtain the main beam temperature $T_{\rm MB}$.  
We employed the position-switching mode, where the on-source integration time of each scan 
was set to be 20 seconds for all the observations. 
The observed position is $\alpha_\mathrm{J2000}=00^h 20^m 27.9^s$, 
$\delta_\mathrm{J2000}=59^{\circ} 17' 01.0''$, 
where the $^{12}$CO ($J=1-0$) line intensity 
is the strongest in IC10 according to \citet{leroy2006molecular}.  
The position is also confirmed to be bright in infrared 
\citep[\emph{JHK}, \emph{Spitzer}/IRAC and \emph{Spitzer}/MIPS,][]{lebouteiller2012oxygen}.  
The off-source position is $3'$ away in azimuth from the on-source position.
The total observation time was 55 hours ($\sim18$ hours for on-source).  
A typical rms noise temperature for each line in the $T_{\rm MB}$ scale is 
$2.5-8.0$ mK for $84-112$ GHz and $31-53$ mK for $112-116$ GHz 
at a frequency resolution of 976.5 kHz. 

\section{Results}
\begin{figure}
\includegraphics[width=1.65\hsize]{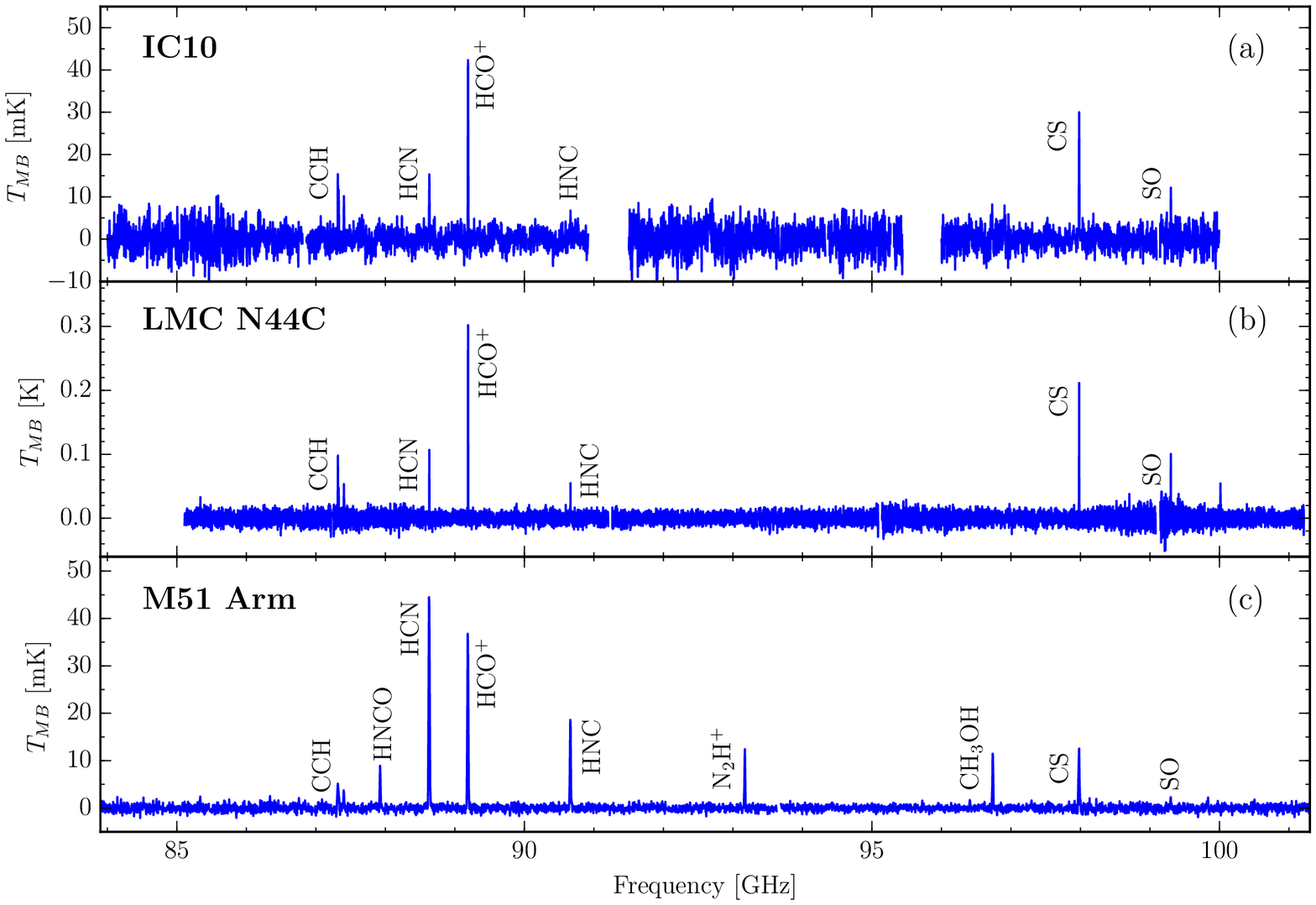}
\caption{Compressed spectra observed toward (a) IC10, 
(b) LMC N44C \citep{nishimura2016spectral}, 
(c) the spiral arm of M51 \citep{watanabe2014spectral}.  
Note that the vertical scale is different from source to source.  
\label{spectra}}
\end{figure}

\begin{figure}
\includegraphics[width=1.75\hsize]{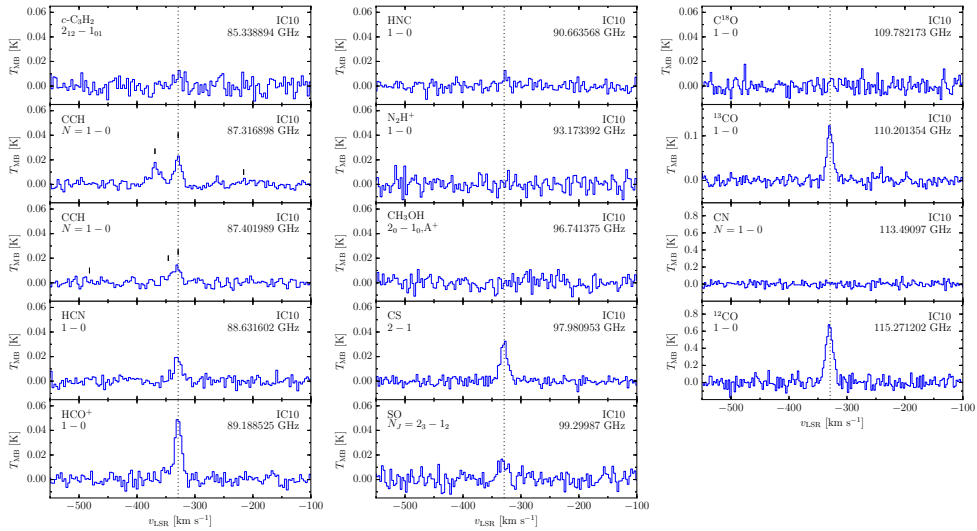}
\caption{Spectral line profiles of individual molecular transitions observed in IC10.  
Small vertical lines in the CCH panel represent 
the positions of the hyperfine components. 
\label{eachline}}
\end{figure}

\begin{figure}
\begin{tabular}{cc}
\includegraphics[width=\hsize]{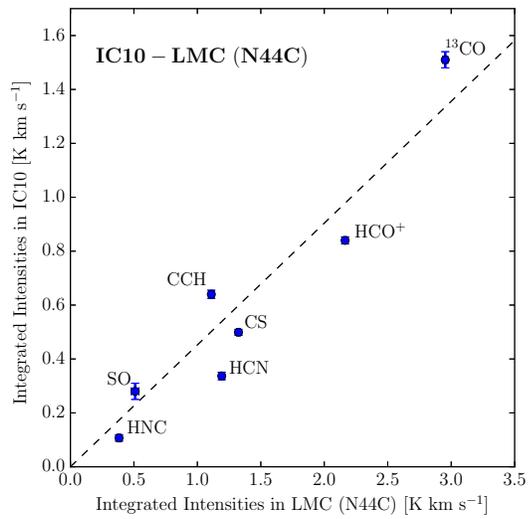}
&
\includegraphics[width=\hsize]{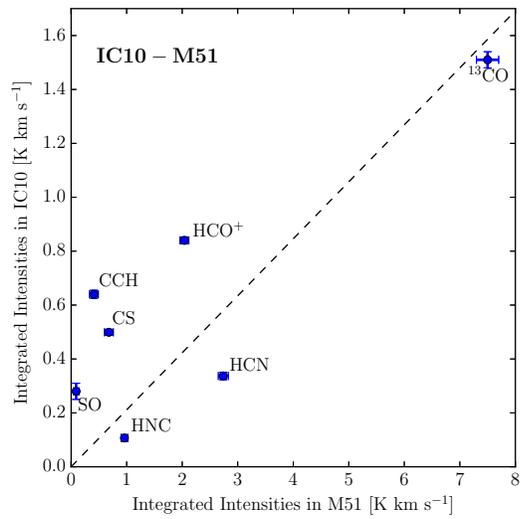}
\end{tabular}
\caption{Correlation diagrams of integrated intensities of detected species 
between IC10 and LMC (N44C) \citep{nishimura2016spectral} (\emph{left}), 
and between IC10 and M51 \citep{watanabe2014spectral} (\emph{right}).  
The dashed line indicates the average ratio of the integrated intensities between the two sources.
\label{correlation_diagram}}
\end{figure}

Figure \ref{spectra} shows the observed spectrum 
in the frequency range from 84 to 100 GHz, 
which is a part of the total observed spectrum.   
We detected the lines of 
CCH($N=1-0$), HCN($J=1-0$), HCO$^+$($J=1-0$), HNC($J=1-0$; tentative detection), 
CS($J=2-1$), and SO($N_J=2_3-1_2$) in this frequency range for the first time.  
In addition, we also detected the lines of $^{13}$CO($J=1-0$) and $^{12}$CO($J=1-0$).  
On the other hand, the lines of $c$-C$_3$H$_2$($2_{12}-1_{01}$), 
N$_2$H$^+$($J=1-0$), CH$_3$OH($J_k=2_0-1_0$, A$^+$), C$^{18}$O($J=1-0$), 
and CN($N=1-0$) were not detected (Figure \ref{eachline}). 
Surprisingly, the spectral intensity pattern is 
very similar to that of the molecular cloud N44C in the LMC, 
although the molecular line intensities are weaker by a factor of about 1/6 
than that toward N44C.  
Figure \ref{correlation_diagram} (left) shows 
the integrated intensities of IC10 versus those of the LMC cloud N44C.  
Indeed a good correlation is seen between the two sources.  
The correlation coefficient is as high as 0.96. 
Even if the $^{13}$CO data are excluded, it is 0.90. 
On the other hand, the spectrum toward IC10 significantly differs 
from that observed in the spiral arm of M51.  
This is also confirmed by the rather poor correlation 
in Figure \ref{correlation_diagram} (right).  
The correlation coefficient is 0.83, while it is as low as 0.18 without $^{13}$CO. 
As mentioned in Section 1, the spectrum of N44C reflects the GMC-scale chemical composition 
characteristic to the LMC, and hence, the similarity 
seems to originate from the low-metallicity environment of IC10.  

The line parameters derived by Gaussian fitting are summarized in Table \ref{lineparameter}.  
The velocity of the peak of the $^{12}$CO line ($-330.2$ km s$^{-1}$) is consistent 
with the previous observation of the $^{12}$CO ($J=1-0$) line toward the same position 
with the Arizona Radio Observatory 12 m telescope (HPBW of $55''$) 
by \citet{leroy2006molecular} ($-330.5$ km s$^{-1}$).  
The integrated intensity of $^{12}$CO observed by \citet{leroy2006molecular} is 
lower than ours only by a factor of 1.3, 
in spite of a large difference of the telescope beam size ($15.3''$ and $55.5''$). 
This result indicates that the $^{12}$CO emitting region is at least extended over 
the beam size of the Nobeyama 45 m telescope.  
The $v_{\rm LSR}$ values of the detected lines range from $-326$ to $-333$ km s$^{-1}$, 
which are consistent with that of the $^{13}$CO line. 
The line widths are mostly in the range from 12 km s$^{-1}$ to 16 km s$^{-1}$.  
Exceptions are the HNC line and one of the CCH lines (87.401989 GHz) 
probably due to a poor signal to noise ratio (3.8$\sigma$) 
and the blending of nearby hyperfine components, respectively (Figure \ref{eachline}).  

We evaluated beam-averaged column densities by statistical equilibrium calculations, 
as we did for the LMC clouds \citep{nishimura2016spectral}. 
We employed the RADEX code \citep{vandertak2007computer} for this purpose.  
Since only one rotational transition was observed for each molecular species, 
we assumed a range of the gas kinetic temperature to be from 10 K to 50 K, 
and a range of the H$_2$ density from $3\times10^3$ cm$^{-3}$ to $1\times10^6$ cm$^{-3}$.  
Althogh the H$_2$ density of $3\times10^5$ cm$^{-3}$ and $10^6$ cm$^{-3}$ are 
seem too high for the H$_2$ density averaged over the 80 pc scale of molecular clouds, 
we assumed the wide range of the physical condition for robust estimate. 
The beam-averaged column densities are derived for 
the gas kinetic temperatures of 10 K, 20 K, 30 K, 40 K, and 50 K 
and the H$_2$ density of $3\times10^3$ cm$^{-3}$, $1\times10^4$ cm$^{-3}$, 
$3\times10^4$ cm$^{-3}$, $1\times10^5$ cm$^{-3}$, $3\times10^5$ cm$^{-3}$, 
and $1\times10^6$ cm$^{-3}$, as listed in Table \ref{columndensity}. 

\section{Discussion}

\subsection{Effect of elemental abundances}
Elemental abundances are different from galaxies to galaxies, 
reflecting the past history of star formation.  
In the low-metallicity galaxies, heavy elements are generally deficient, 
and in particular, the deficiency of nitrogen is the most significant 
among the abundant second row elements (C, N, O) \citep{vincenzo2016nitrogen}. 
Because of this reason, one of the characteristic features in the chemical compositions 
would be the deficiency of the N-bearing molecules.  
The chemical model by \citet{millar1990chemical} indeed predicted that 
abundances of N-bearing species are sensitive to the elemental abundance of nitrogen.  
The deficiency of the N-bearing molecules is already evident 
in the spectral pattern of IC10 in comparison with that of the spiral arm of M51.  
We calculated the abundance ratios of HCN/HCO$^+$ and HNC/HCO$^+$ for IC10, 
under the assumption of the H$_2$ density of $3\times10^3$, 
$1\times10^4$, $3\times10^4$, and $1\times10^5$ cm$^{-3}$, 
$3\times10^5$, and $1\times10^6$ cm$^{-3}$, 
and the gas kinetic temperature of 10, 20, 30, 40, and 50 K, 
and compared them with the average ratios reported for the seven clouds in the LMC \citep{nishimura2016spectral}, 
the average ratios for three translucent clouds (CB17, CB24, CB228) 
in our Galaxy \citep{turner1995physicsB, turner1997physics}, 
and the ratio for the spiral arm position (P1) of M51 
observed by \citet{watanabe2014spectral}, as shown in Table \ref{ratio}.  
The observed spectrum of IC10 are averaged over molecular clouds ($\sim80$ pc scale), 
and seems to be dominated by the diffuse part of molecular clouds 
rather than dense star-forming cores. 
This is true even for smaller-scale observations 
($\sim10$ pc scale) of the LMC clouds \citep{nishimura2016spectral}. 
Hence, for fair comparison, we chose the translucent clouds as representatives of our Galaxy.  
Although the column densities are sensitive to the assumed H$_2$ density 
and the assumed gas kinetic temperature, the column density ratios for IC10 
are affected only less than $\pm50$\% in the parameter range.  
This fact is also shown in the analyses of the LMC spectra \citep{nishimura2016spectral}.
The elemental N/O ratio in IC10 is lower by a factor of 3 than that in our Galaxy 
\citep{lequeux1979chemical}.  
The HCN/HCO$^+$ and HNC/HCO$^+$ ratios in IC10 are comparable to those for the LMC clouds, and 
are indeed found to be lower than in the three Galactic translucent clouds (CB17, CB24, CB228).  
Although the HCN/HCO$^+$ ratios of IC10 and the Galactic translucent clouds marginally overlap 
with each other within the mutual error ranges, 
the overlap occurs only for the particular conditions that is not very likely 
($3\times10^3$ cm$^{-3}$ and 10 K for IC10 and 
$1\times10^5$ cm$^{-3}$ and 50 K for the Galactic translucent clouds). 
Hence, we can state the above conclution in spite of the formal error ranges. 
A similar comparison can also be made for M51 P1, 
which has the higher N/O ratio than the Solar neighborhood by a factor of 2 
\citep{bresolin2004abundance}. 
The HCN/HCO$^+$ and HNC/HCO$^+$ ratios in M51 P1 are $8.4^{+4.0}_{-4.6}$ 
and $1.6^{+0.6}_{-0.6}$, respectively, which are higher than those in IC10 
($1.9^{+1.9}_{-1.5}$ and $0.4^{+0.2}_{-0.2}$, respectively).  
Hence, it is most likely that the deficiency of the N-bearing molecules 
directly reflects the elemental deficiency of nitrogen in the IC10.  

The HCN/HCO$^+$ ratio has been discussed for nuclear regions of external galaxies. 
The HCN/HCO$^+$ ratio is known to be higher for the AGNs, 
which is interpreted as the effect of XDRs, cosmic-rays, and/or shock heatings 
\citep[e.g.,][]{lepp1996xray, kohno2001dense, meijerink2007diagnostic, aladro2015lambda}. 
In this study, we observed GMCs without such effects, 
and found the lower HCN/HCO$^+$ ratio than the Galactic translucent clouds. 
Hence, the various effects suggested for the AGNs cannot be applied to IC10. 
Rather, the intrinsic effect of the elemental abundance can be seen in this source. 

The CS/SO ratio in IC10 is $0.9^{+0.5}_{-0.5}$, which is comparable to 
that of the LMC ($1.8^{+0.4}_{-0.3}$).  
The ratio is also comparable to that in the Galactic translucent clouds 
($1.0^{+0.4}_{-0.5}$), but is lower than that of M51 P1 ($4.6^{+1.2}_{-1.8}$).  
Any significant trend due to the difference of the C/O ratio 
is not seen in the CS/SO ratio in this study.  

\subsection{Effect of photodissociation}
We calculated the abundance ratio of CCH/HCO$^+$ in the same way as HCN/HCO$^+$ and HNC/HCO$^+$.  
The ratio of CCH/HCO$^+$ is higher in IC10 ($20.9^{+10.7}_{-9.2}$) 
than in the Galactic translucent clouds ($5.3^{+3.9}_{-2.4}$) by a factor of 4.  
This enhancement of CCH in IC10 cannot be interpreted 
as the effect of elemental abundances. 
Indeed, the elemental C/O ratio is estimated to be 0.3 in IC10 
\citep{lequeux1979chemical, bolatto2000submillimeter}, 
while it is 0.6 in Solar neighborhood.  
When the low C/O ratio is taken into account,
the high CCH/HCO$^+$ ratio in IC10 is striking.  
This enhancement of CCH is also seen in the LMC clouds \citep{nishimura2016spectral}. 

The effect of photodissociation would be responsible for the enhancement. 
It is generally thought that CCH is abundant in the photodissociation region (PDR) 
illuminated by UV radiation 
\citep[e.g.,][]{pety2005are, martin2014chemistry, ginard2015chemical}.  
In low-metallicity galaxies, the extinction of the UV radiation by dust grains 
is expected to be less effective for a given column density of H$_2$ 
because of the lower abundance of dust grains.  
The PDR would be extended deeper into molecular clouds, 
which would be responsible to the relatively high abundance of CCH.  
In the PDR, the growth of large carbon-chain molecules 
containing more than three carbon atoms are generally suppressed 
by competitive photodissociation processes \citep[e.g.,][]{lucas2000comparative}. 
The $c$-C$_3$H$_2$/CCH ratio is indeed found to be less than 0.1 both in IC10 the LMC clouds, 
which is lower than that observed in the Galactic translucent clouds 
\citep[0.22;][]{turner1999physics, turner2000physics}.  
The ratio is rather consistent with the ratio in some Galactic diffuse clouds 
observed in absorption against the bright continuum sources \citep[0.04;][]{lucas2000comparative} 
and the ratio in M82 which also hosts extended PDRs 
\citep[0.04;][]{fuente2005photon, aladro2015lambda}. 
This fact further supports the extended PDR in the low-metallicity galaxies.  

Non-detection of CH$_3$OH is notable, 
as in the case of the LMC clouds \citep{nishimura2016spectral}.  
We obtained the upper limits of the CH$_3$OH intensity and the column density in IC10.  
The abundance ratio of CH$_3$OH/HCO$^+$ in IC10 is $<2.2$, 
which seems lower than that found in M51 ($3.8^{+5.8}_{-2.3}$).  
This result can also be interpreted in terms of a stronger UV effect 
owing to low abundance of dust grains in IC10. 
CH$_3$OH is thought to be produced by hydrogenation of CO 
on dust grains, and is liberated into the gas phase by thermal 
and/or non-thermal desorption \citep[e.g.,][]{watanabe2002efficient}.
A low abundance of dust grains tends to make the CH$_3$OH formation inefficient. 
Furthermore, laboratory experiments show that 
the efficiency of CH$_3$OH formation significantly decreases 
at the temperature higher than 20 K due to a fall of striking probability of hydrogen atom
\citep{watanabe2003dependence}.  
Since the temperature of cloud peripheries 
is expected to be higher in the low-metallicity condition 
due to penetration of the UV radiation, CH$_3$OH would not be formed efficiently. 
According to \citet{shimonishi2016vlt}, 
the lower abundance of CH$_3$OH ice observed in the LMC may be also caused by 
a relatively high dust temperature.  
Their result is consistent with ours.  

We also evaluated the HNC/HCN ratio in IC10 to be $0.22^{+0.12}_{-0.11}$, 
which is comparable to the ratio in the LMC clouds, 
and is lower than the ratio in typical dark clouds \citep[$0.54-4.5$;][]{hirota1998abundances}
in the Solar neighborhood.  
It is close to the ratio reported in some Galactic diffuse clouds, where the HCN and HNC lines 
are detected in absorption against bright continuum sources 
\citep[$0.21\pm0.05$;][]{liszt2001comparative}.  
\citet{hirota1998abundances} reported that the HNC/HCN ratio 
is lower under higher-temperature environments: the ratio decreases above 24 K, 
possibly reflecting isomerization mechanisms of HNC to HCN. 
The relatively low ratios observed in IC10 and the LMC clouds 
may also originate from warmer temperature conditions 
due to higher UV field and/or lower grain abundance.  
In addition, it is worth noting non-detection of N$_2$H$^+$.  
This seems to originate mainly from the low elemental abundance of nitrogen.  
In addition, the UV radiation may also contribute to the low-abundance of N$_2$H$^+$. 
Deeper penetration of UV radiation enhances the abundance of atomic ion and electrons, 
which efficiently destroys N$_2$H$^+$ \citep{aikawa2015analytical}.  

\section{Summary}

The molecular-cloud-scale chemical composition of IC10 is found to be 
very similar to that of the LMC clouds.  
It is characterized by deficiency of N-bearing molecules, 
relatively high abundant CCH, and deficiency of CH$_3$OH.  
Hence, these chemical features can be regarded as ones characteristic in low-metallicity galaxies, 
although they have to be further examined in other galaxies with various metallicities.  
Furthermore, more sensitive observations are needed 
to detect larger molecules and explore the molecular evolution in low-metallicity galaxies.  

\acknowledgments
We thank an anonymous reviewer for valuable coments. 
We thank the staff of the NRO 45 m telescope for excellent support.  
This study is partly supported from Grants-in-Aid of Education, Sports, 
Science, and Technologies of Japan (21224002, 25400223, and 25108005).  
YN is supported by Grant-in-Aid for JSPS Fellows (268280).

\clearpage

%%%%%%%%%%%%%%%%%%%%%%%%%%%%%%%%%%%%%%%%%%%%%%%%%%

\clearpage

%%%%%%%%%%%%%%%%%%%%%%%%%%%%%%%%%%%%%%%%%%%%%%%%%%

\begin{deluxetable}{lrlrrrrr}
\tabletypesize{\scriptsize}
\rotate
\tablecaption{IC10 Observed Line Parameters \label{lineparameter}}
\tablewidth{0pt}
\tablehead{
\colhead{Molecule} & \colhead{Frequency} 
& \colhead{Transition} & \colhead{$T_{\rm MB}$ Peak} 
& \colhead{$v_{\rm LSR}$} & \colhead{$\Delta v$}
& \colhead{$\int T_{\rm MB}dv$} \\
\colhead{} & \colhead{(GHz)} 
& \colhead{} & \colhead{(mK)} & \colhead{(km s$^{-1}$)} 
& \colhead{(km s$^{-1}$)} & \colhead{(K km s$^{-1}$)}}
\startdata
$c$-C$_3$H$_2$ & 85.338894 & $2_{12}-1_{01}$ &
&
&
&
$ < 0.12 $ \\
CCH & 87.284105 & $N=1-0$, $J=5/2-3/2$, $F=1-1$ &
&
&
&
$ < 0.05 $ \\
CCH & 87.316898 & $N=1-0$, $J=5/2-3/2$, $F=2-1$ &
$ 20 \pm 2 $ &
$ -328.9 \pm 0.6 $ &
$ 15.5 \pm 1.4 $ &
$ 0.329 \pm 0.015 $ \\
CCH & 87.328585 & $N=1-0$, $J=5/2-3/2$, $F=1-0$ &
$ 15.7 \pm 1.9 $ &
$-$ &
$-$ &
$ 0.311 \pm 0.014 $ \\
CCH & 87.401989 & $N=1-0$, $J=3/2-3/2$, $F=1-1$ &
$  12 \pm   3 $ &
$ -333 \pm   2 $ &
$  21 \pm   5 $ &
$ 0.28 \pm 0.03 $ \\
CCH & 87.407165 & $N=1-0$, $J=3/2-3/2$, $F=0-1$ &
$-$ &
$-$ &
$-$ &
$-$ \\
CCH & 87.446470 & $N=1-0$, $J=3/2-3/2$, $F=1-0$ &
&
&
&
$ < 0.05 $ \\
HCN & 88.631602 & $1-0$ &
$  19 \pm   2 $ &
$ -329.8 \pm 0.8 $ &
$ 15 \pm 2 $ &
$ 0.337 \pm 0.014 $ \\
HCO$^+$ & 89.188525 & $1-0$ &
$  49 \pm   2 $ &
$ -328.9 \pm 0.3 $ &
$ 13.9 \pm 0.8 $ &
$ 0.840 \pm 0.013 $ \\
HNC & 90.663568 & $1-0$ &
$  10 \pm   3 $ &
$ -326.7 \pm 1.2 $ &
$   9 \pm   3 $ &
$ 0.107 \pm 0.012 $ \\
N$_2$H$^+$ & 93.173392 & $1-0$ &
&
&
&
$ < 0.11 $ \\
CH$_3$OH & 96.741375 & $2_0-1_0$, A$^+$ &
&
&
&
$ < 0.09 $ \\
CS & 97.980953 & $2-1$ &
$ 30.7 \pm 1.6 $ &
$ -329.2 \pm 0.4 $ &
$ 15.0 \pm 0.9 $ &
$ 0.499 \pm 0.011 $ \\
SO & 99.299870 & $N_J=2_3-1_2$ &
$  14 \pm   3 $ &
$ -331.2 \pm 1.7 $ &
$  18 \pm   4 $ &
$ 0.28 \pm 0.03 $ \\
C$^{18}$O & 109.782173 & $1-0$ &
&
&
&
$ < 0.10 $ \\
$^{13}$CO & 110.201354 & $1-0$ &
$ 118 \pm   5 $ &
$ -329.4 \pm 0.3 $ &
$ 12.0 \pm 0.6 $ &
$ 1.51 \pm 0.03 $ \\
CN & 113.490970 & $N=1-0$, $J=3/2-1/2$, $F=5/2-3/2$ &
&
&
&
$ < 0.6 $ \\
$^{12}$CO & 115.271202 & $1-0$ &
$ 653 \pm  31 $ &
$ -330.2 \pm 0.3 $ &
$ 15.1 \pm 0.8 $ &
$ 10.4 \pm 0.2 $
\enddata
\tablecomments{The errors are 1$\sigma$. The upper limits are 3$\sigma$. 
The calibration error ($\sim20$\%) is not included.}
\end{deluxetable}

\begin{deluxetable}{lcccccc}
\tabletypesize{\scriptsize}
\tablecaption{Derived column densities\tablenotemark{a}
\label{columndensity}}
\tablewidth{0pt}
\tablecolumns{8}
\tablehead{\\[-7.5mm]}
\startdata
& $n_{\rm H_2}$
& \multicolumn{5}{c}{$3\times10^{3}$ cm$^{-3}$} \\[1mm]
\cline{3-7}\\[-1.5mm]
molecule & $T_k$
& 10 K & 20 K & 30 K & 40 K & 50 K \\[1mm]
\hline\\[-1.5mm]
$c$-C$_3$H$_2$ (ortho) & &
$<2.7(+13)$ & $<1.1(+13)$ & $<6.7(+12)$ & $<5.5(+12)$ & $<5.2(+12)$ \\
CCH & &
$2.4(+14)$ & $1.1(+14)$ & $8.6(+13)$ & $7.1(+13)$ & $6.2(+13)$ \\
HCN & &
$5.3(+13)$ & $2.8(+13)$ & $2.2(+13)$ & $1.9(+13)$ & $1.6(+13)$ \\
HCO$^+$ & &
$1.4(+13)$ & $8.3(+12)$ & $6.7(+12)$ & $5.8(+12)$ & $5.3(+12)$ \\
HNC & &
$6.1(+12)$ & $4.1(+12)$ & $3.6(+12)$ & $3.3(+12)$ & $3.1(+12)$ \\
N$_2$H$^+$ & &
$<1.8(+12)$ & $<1.1(+12)$ & $<9.2(+11)$ & $<8.0(+11)$ & $<7.3(+11)$ \\
CH$_3$OH (A) & &
$<8.7(+12)$ & $<4.9(+12)$ & $<4.0(+12)$ & $<3.6(+12)$ & $<3.4(+12)$ \\
CS & &
$6.5(+13)$ & $3.8(+13)$ & $3.0(+13)$ & $2.5(+13)$ & $2.3(+13)$ \\
SO & &
$6.0(+13)$ & $2.8(+13)$ & $2.1(+13)$ & $1.8(+13)$ & $1.6(+13)$ \\
C$^{18}$O & &
$<8.5(+13)$ & $<7.8(+13)$ & $<8.1(+13)$ & $<8.5(+13)$ & $<8.9(+13)$ \\
$^{13}$CO & &
$1.3(+15)$ & $1.2(+15)$ & $1.2(+15)$ & $1.3(+15)$ & $1.3(+15)$ \\
CN & &
$<5.0(+14)$ & $<2.2(+14)$ & $<1.5(+14)$ & $<1.1(+14)$ & $<9.4(+13)$ \\
$^{12}$CO & &
$9.2(+15)$ & $7.8(+15)$ & $7.9(+15)$ & $8.2(+15)$ & $8.6(+15)$ \\[1mm]
%%%
\hline
\\
& $n_{\rm H_2}$
& \multicolumn{5}{c}{$1\times10^{4}$ cm$^{-3}$} \\[1mm]
\cline{3-7}\\[-1.5mm]
molecule & $T_k$
& 10 K & 20 K & 30 K & 40 K & 50 K \\[1mm]
\hline\\[-1.5mm]
$c$-C$_3$H$_2$ (ortho) & &
$<7.9(+12)$ & $<3.3(+12)$ & $<2.1(+12)$ & $<1.8(+12)$ & $<1.7(+12)$ \\
CCH & &
$7.8(+13)$ & $4.1(+13)$ & $3.2(+13)$ & $2.8(+13)$ & $2.5(+13)$ \\
HCN & &
$1.4(+13)$ & $8.2(+12)$ & $6.4(+12)$ & $5.5(+12)$ & $4.9(+12)$ \\
HCO$^+$ & &
$4.3(+12)$ & $2.7(+12)$ & $2.2(+12)$ & $1.9(+12)$ & $1.8(+12)$ \\
HNC & &
$1.8(+12)$ & $1.3(+12)$ & $1.1(+12)$ & $1.0(+12)$ & $9.9(+11)$ \\
N$_2$H$^+$ & &
$<6.0(+11)$ & $<3.8(+11)$ & $<3.2(+11)$ & $<2.8(+11)$ & $<2.6(+11)$ \\
CH$_3$OH (A) & &
$<3.5(+12)$ & $<2.3(+12)$ & $<2.0(+12)$ & $<1.9(+12)$ & $<1.8(+12)$ \\
CS & &
$2.0(+13)$ & $1.2(+13)$ & $9.5(+12)$ & $8.2(+12)$ & $7.4(+12)$ \\
SO & &
$1.9(+13)$ & $9.4(+12)$ & $7.2(+12)$ & $6.3(+12)$ & $5.7(+12)$ \\
C$^{18}$O & &
$<8.6(+13)$ & $<9.6(+13)$ & $<1.1(+14)$ & $<1.3(+14)$ & $<1.4(+14)$ \\
$^{13}$CO & &
$1.3(+15)$ & $1.5(+15)$ & $1.7(+15)$ & $1.9(+15)$ & $2.1(+15)$ \\
CN & &
$<1.3(+14)$ & $<6.0(+13)$ & $<4.2(+13)$ & $<3.4(+13)$ & $<2.9(+13)$ \\
$^{12}$CO & &
$8.9(+15)$ & $9.4(+15)$ & $1.1(+16)$ & $1.2(+16)$ & $1.3(+16)$ \\[1mm]
%%%
\hline
\\
& $n_{\rm H_2}$
& \multicolumn{5}{c}{$3\times10^{4}$ cm$^{-3}$} \\[1mm]
\cline{3-7}\\[-1.5mm]
molecule & $T_k$
& 10 K & 20 K & 30 K & 40 K & 50 K \\[1mm]
\hline\\[-1.5mm]
$c$-C$_3$H$_2$ (ortho) & &
$<2.8(+12)$ & $<1.3(+12)$ & $<8.7(+11)$ & $<7.6(+11)$ & $<7.3(+11)$ \\
CCH & &
$3.4(+13)$ & $2.1(+13)$ & $1.8(+13)$ & $1.6(+13)$ & $1.6(+13)$ \\
HCN & &
$4.9(+12)$ & $2.8(+12)$ & $2.3(+12)$ & $2.0(+12)$ & $1.8(+12)$ \\
HCO$^+$ & &
$1.7(+12)$ & $1.2(+12)$ & $1.0(+12)$ & $9.3(+11)$ & $8.8(+11)$ \\
HNC & &
$6.9(+11)$ & $4.9(+11)$ & $4.3(+11)$ & $4.1(+11)$ & $3.9(+11)$ \\
N$_2$H$^+$ & &
$<2.6(+11)$ & $<1.8(+11)$ & $<1.5(+11)$ & $<1.4(+11)$ & $<1.4(+11)$ \\
CH$_3$OH (A) & &
$<2.2(+12)$ & $<1.7(+12)$ & $<1.7(+12)$ & $<1.7(+12)$ & $<1.8(+12)$ \\
CS & &
$7.5(+12)$ & $4.7(+12)$ & $3.9(+12)$ & $3.5(+12)$ & $3.2(+12)$ \\
SO & &
$7.4(+12)$ & $4.2(+12)$ & $3.5(+12)$ & $3.2(+12)$ & $3.1(+12)$ \\
C$^{18}$O & &
$<9.3(+13)$ & $<1.2(+14)$ & $<1.4(+14)$ & $<1.7(+14)$ & $<1.9(+14)$ \\
$^{13}$CO & &
$1.4(+15)$ & $1.7(+15)$ & $2.1(+15)$ & $2.5(+15)$ & $2.9(+15)$ \\
CN & &
$<4.1(+13)$ & $<2.1(+13)$ & $<1.5(+13)$ & $<1.3(+13)$ & $<1.1(+13)$ \\
$^{12}$CO & &
$9.5(+15)$ & $1.1(+16)$ & $1.4(+16)$ & $1.6(+16)$ & $1.8(+16)$ \\[1mm]
%%%
\hline
\\[5mm]
& $n_{\rm H_2}$
& \multicolumn{5}{c}{$1\times10^{5}$ cm$^{-3}$} \\[1mm]
\cline{3-7}\\[-1.5mm]
molecule & $T_k$
& 10 K & 20 K & 30 K & 40 K & 50 K \\[1mm]
\hline\\[-1.5mm]
$c$-C$_3$H$_2$ (ortho) & &
$<1.0(+12)$ & $<5.6(+11)$ & $<4.7(+11)$ & $<4.5(+11)$ & $<4.4(+11)$ \\
CCH & &
$1.9(+13)$ & $1.5(+13)$ & $1.5(+13)$ & $1.5(+13)$ & $1.6(+13)$ \\
HCN & &
$1.7(+12)$ & $1.0(+12)$ & $8.7(+11)$ & $7.8(+11)$ & $7.2(+11)$ \\
HCO$^+$ & &
$8.9(+11)$ & $6.9(+11)$ & $6.5(+11)$ & $6.3(+11)$ & $6.3(+11)$ \\
HNC & &
$2.9(+11)$ & $2.2(+11)$ & $2.0(+11)$ & $1.9(+11)$ & $1.8(+11)$ \\
N$_2$H$^+$ & &
$<1.4(+11)$ & $<1.1(+11)$ & $<1.0(+11)$ & $<1.0(+11)$ & $<1.0(+11)$ \\
CH$_3$OH (A) & &
$<1.9(+12)$ & $<1.9(+12)$ & $<2.1(+12)$ & $<2.3(+12)$ & $<2.5(+12)$ \\
CS & &
$3.3(+12)$ & $2.3(+12)$ & $2.1(+12)$ & $2.0(+12)$ & $1.9(+12)$ \\
SO & &
$3.8(+12)$ & $2.8(+12)$ & $2.7(+12)$ & $2.7(+12)$ & $2.8(+12)$ \\
C$^{18}$O & &
$<9.7(+13)$ & $<1.3(+14)$ & $<1.6(+14)$ & $<2.0(+14)$ & $<2.3(+14)$ \\
$^{13}$CO & &
$1.5(+15)$ & $1.9(+15)$ & $2.5(+15)$ & $3.0(+15)$ & $3.5(+15)$ \\
CN & &
$<1.4(+13)$ & $<7.9(+12)$ & $<6.2(+12)$ & $<5.4(+12)$ & $<4.9(+12)$ \\
$^{12}$CO & &
$9.9(+15)$ & $1.2(+16)$ & $1.6(+16)$ & $1.9(+16)$ & $2.2(+16)$ \\[1mm]
%%%
\hline
\\
& $n_{\rm H_2}$
& \multicolumn{5}{c}{$3\times10^{5}$ cm$^{-3}$} \\[1mm]
\cline{3-7}\\[-1.5mm]
molecule & $T_k$
& 10 K & 20 K & 30 K & 40 K & 50 K \\[1mm]
\hline\\[-1.5mm]
$c$-C$_3$H$_2$ (ortho) & &
$<5.5(+11)$ & $<4.3(+11)$ & $<4.8(+11)$ & $<5.2(+11)$ & $<5.4(+11)$ \\
CCH & &
$1.7(+13)$ & $1.7(+13)$ & $1.9(+13)$ & $2.1(+13)$ & $2.4(+13)$ \\
HCN & &
$7.8(+11)$ & $5.5(+11)$ & $4.9(+11)$ & $4.6(+11)$ & $4.4(+11)$ \\
HCO$^+$ & &
$7.1(+11)$ & $6.6(+11)$ & $6.8(+11)$ & $7.2(+11)$ & $7.5(+11)$ \\
HNC & &
$1.8(+11)$ & $1.5(+11)$ & $1.4(+11)$ & $1.4(+11)$ & $1.3(+11)$ \\
N$_2$H$^+$ & &
$<1.2(+11)$ & $<1.1(+11)$ & $<1.1(+11)$ & $<1.2(+11)$ & $<1.3(+11)$ \\
CH$_3$OH (A) & &
$<2.0(+12)$ & $<2.2(+12)$ & $<2.7(+12)$ & $<3.1(+12)$ & $<3.5(+12)$ \\
CS & &
$2.2(+12)$ & $1.8(+12)$ & $1.8(+12)$ & $1.9(+12)$ & $1.9(+12)$ \\
SO & &
$3.1(+12)$ & $3.0(+12)$ & $3.4(+12)$ & $3.8(+12)$ & $4.1(+12)$ \\
C$^{18}$O & &
$<9.8(+13)$ & $<1.3(+14)$ & $<1.7(+14)$ & $<2.1(+14)$ & $<2.6(+14)$ \\
$^{13}$CO & &
$1.5(+15)$ & $2.0(+15)$ & $2.6(+15)$ & $3.2(+15)$ & $3.9(+15)$ \\
CN & &
$<6.6(+12)$ & $<4.4(+12)$ & $<3.8(+12)$ & $<3.6(+12)$ & $<3.6(+12)$ \\
$^{12}$CO & &
$1.0(+16)$ & $1.3(+16)$ & $1.7(+16)$ & $2.0(+16)$ & $2.4(+16)$ \\[1mm]
%%%
\hline
\\
& $n_{\rm H_2}$
& \multicolumn{5}{c}{$1\times10^{6}$ cm$^{-3}$} \\[1mm]
\cline{3-7}\\[-1.5mm]
molecule & $T_k$
& 10 K & 20 K & 30 K & 40 K & 50 K \\[1mm]
\hline\\[-1.5mm]
$c$-C$_3$H$_2$ (ortho) & &
$<4.5(+11)$ & $<5.8(+11)$ & $<8.7(+11)$ & $<1.1(+12)$ & $<1.2(+12)$ \\
CCH & &
$1.8(+13)$ & $2.3(+13)$ & $2.8(+13)$ & $3.3(+13)$ & $3.8(+13)$ \\
HCN & &
$5.0(+11)$ & $4.3(+11)$ & $4.3(+11)$ & $4.4(+11)$ & $4.5(+11)$ \\
HCO$^+$ & &
$7.4(+11)$ & $8.5(+11)$ & $1.0(+12)$ & $1.1(+12)$ & $1.3(+12)$ \\
HNC & &
$1.5(+11)$ & $1.4(+11)$ & $1.4(+11)$ & $1.5(+11)$ & $1.5(+11)$ \\
N$_2$H$^+$ & &
$<1.2(+11)$ & $<1.4(+11)$ & $<1.7(+11)$ & $<2.0(+11)$ & $<2.2(+11)$ \\
CH$_3$OH (A) & &
$<2.1(+12)$ & $<2.6(+12)$ & $<3.3(+12)$ & $<4.0(+12)$ & $<4.7(+12)$ \\
CS & &
$2.1(+12)$ & $2.1(+12)$ & $2.3(+12)$ & $2.6(+12)$ & $2.9(+12)$ \\
SO & &
$3.2(+12)$ & $3.8(+12)$ & $4.8(+12)$ & $5.7(+12)$ & $6.5(+12)$ \\
C$^{18}$O & &
$<9.9(+13)$ & $<1.3(+14)$ & $<1.8(+14)$ & $<2.2(+14)$ & $<2.7(+14)$ \\
$^{13}$CO & &
$1.5(+15)$ & $2.0(+15)$ & $2.7(+15)$ & $3.4(+15)$ & $4.0(+15)$ \\
CN & &
$<4.2(+12)$ & $<3.6(+12)$ & $<3.7(+12)$ & $<3.9(+12)$ & $<4.2(+12)$ \\
$^{12}$CO & &
$1.0(+16)$ & $1.3(+16)$ & $1.7(+16)$ & $2.1(+16)$ & $2.5(+16)$ 
\enddata
\tablenotetext{a}{$a(+b)$ refers to $a\times10^{+b}$ cm$^{-2}$.}
\end{deluxetable}

\begin{deluxetable}{llcccc}
%\rotate
\tabletypesize{\scriptsize}
\tablecaption{Column density and elemental abundance ratios. \label{ratio}}
\tablewidth{0pt}
\tablecolumns{5}
\tablehead{
\colhead{} & \colhead{} 
& \colhead{IC10} & \colhead{LMC} & \colhead{``Solar''} & \colhead{M51} }
\startdata
Elemental abundance ratio 
& N/O                   &      0.04  & 0.036 & 0.12  & $\sim0.25$ \\
& C/O                   &      0.3   & 0.33  & 0.60  & $\sim0.6$  \\
& S/O                   & $\sim0.03$ & 0.043 & 0.023 & $\sim0.025$  \\[1mm]
\hline\\[-1.5mm]
Column density ratio
& $N$[HCN]/$N$[HCO$^+$] & $1.9^{+1.9}_{-1.5}$    & $4.0^{+0.6}_{-1.0}$ 
                        & $8.0^{+2.9}_{-4.6}$    & $8.4^{+4.0}_{-4.6}$ \\
& $N$[HNC]/$N$[HCO$^+$] & $0.4^{+0.2}_{-0.2}$    & $0.85^{+0.15}_{-0.12}$ 
                        & $3.4^{+1.3}_{-1.2}$    & $1.6^{+0.6}_{-0.6}$ \\
& $N$[HNC]/$N$[HCN]     & $0.22^{+0.12}_{-0.11}$ & $0.23^{+0.06}_{-0.06}$ 
                        & $0.4^{+0.2}_{-0.2}$    & $0.21^{+0.07}_{-0.07}$ \\
& $N$[CCH]/$N$[HCO$^+$] & $20.9^{+10.7}_{-9.2}$  & $12.4^{+3.3}_{-3.1}$ 
                        & $5.3^{+3.9}_{-2.4}$    & $9.1^{+3.8}_{-2.9}$ \\
& $N$[CS] /$N$[HCO$^+$] & $3.5^{+1.1}_{-1.3}$    & $4.6^{+0.3}_{-0.5}$ 
                        & $3.4^{+0.5}_{-0.7}$    & $2.3^{+0.4}_{-0.6}$ \\
& $N$[CS] /$N$[SO]      & $0.9^{+0.5}_{-0.5}$    & $1.8^{+0.4}_{-0.3}$ 
                        & $1.0^{+0.4}_{-0.5}$    & $4.6^{+1.2}_{-1.8}$ \\
\enddata
\tablecomments{
Elemental abundances are based on 
\citet{lequeux1979chemical, bolatto2000submillimeter, magrini2009ic10} for IC10, 
\citet{dufour1982carbon} for the LMC and ``Solar'', 
\citet{bresolin2004abundance, garnett2004CNO} for M51.  
Column densities of the LMC are calculated 
for the H$_2$ density range of $3\times10^3 - 3\times10^4$ cm$^{-3}$ and
the gas kinetic temperature range of $10-30$ K, 
based on the literature data \citep{nishimura2016spectral}. 
Those for ``Solar'' (nearby translucent clouds; CB17, CB24, CB228) 
\citep{turner1995physicsA, turner1995physicsB, turner1996physics, 
turner1997physics, turner2000physics}
are calculated for the H$_2$ density range of $3\times10^3 - 1\times10^5$ cm$^{-3}$ and 
the gas kinetic temperature range of $10-50$ K. 
We choose these clouds, because all molecules listed in this table are observed. 
For M51, we calculated the column densities 
under the assumption of H$_2$ density of $3\times10^3 - 1\times10^5$ cm$^{-3}$ 
and the gas kinetic temperature of $10-50$ K, 
based on the literature data \citep{watanabe2014spectral}. 
Note that the H$_2$ density is derived toward this source to be $\sim10^4$ cm$^{-3}$ 
by observations of the H$_2$CO lines \citep{nishimurainprep}. 
The errors are estimated from the variation 
due to the assumed gas kinetic temperature and H$_2$ density.  }
\end{deluxetable}

\end{document}